# Physics-Informed Neural Network-Based Discovery of Hyperelastic Constitutive Models from Extremely Scarce Data

Hyeonbin Moon [1†], Donggeun Park [1†], Hanbin Cho [1], Hong-Kyun Noh[2], Jae hyuk Lim [2*] and Seunghwa Ryu[1*]

[1]Department of Mechanical Engineering, Korea Advanced Institute of Science and Technology (KAIST), 291 Daehak-ro, Yuseong-gu, Daejeon 34141, Republic of Korea

[2] Department of Mechanical Engineering, Kyung Hee University, 1732 Deogyeong-daero, Giheung-gu, Yongin-si, Gyeonggi-do 17104, Republic of Korea

[†] Hyeonbin Moon and Donggeun Park contributed equally to this work.

[*]Corresponding authors: ryush@kaist.ac.kr (Seunghwa Ryu), jaehyuklim@khu.ac.kr (Jae Hyuk Lim)

**Abstract**

The discovery of constitutive models for hyperelastic materials is essential yet challenging due to their nonlinear behavior and the limited availability of experimental data. Traditional methods typically require extensive stress-strain or full-field measurements, which are often difficult to obtain in practical settings. To overcome these challenges, we propose a physics-informed neural network (PINN)-based framework that enables the discovery of constitutive models using only sparse measurement data – such as displacement and reaction force – that can be acquired from a single material test. By integrating PINNs with finite element discretization, the framework reconstructs full-field displacement and identifies the underlying strain energy density from predefined candidates, while ensuring consistency with physical laws. A two-stage training process is employed: the Adam optimizer jointly updates neural network parameters and model coefficients to obtain an initial solution, followed by L-BFGS refinement and sparse regression with $l_p$ regularization to extract a parsimonious constitutive model. Validation on benchmark hyperelasticmodels demonstrates that the proposed method can accurately recover constitutive laws and displacement fields, even when the input data are limited and noisy. These findings highlight the applicability of the proposed framework to experimental scenarios where measurement data are both scarce and noisy.

# 1. Introduction

Constitutive models define the relationship between stress and strain and are central to engineering analyses and material behavior prediction. Over the past few decades, extensive research has resulted in a variety of mathematical formulations that simulate real-world material responses, encompassing elasticity, plasticity, viscoplasticity, and hyperelasticity [1-4]. However, unlike fundamental physical conservation laws or kinematic relations, identifying constitutive models that satisfy thermodynamic constraints while accurately capturing the nonlinear and complex behavior of materials remains a major challenge. Traditionally, these models have been derived largely from empirical observations and researchers' intuition, making the selection process highly dependent on prior assumptions [5-8]. Recent advancements in computational and data-driven methodologies have introduced new possibilities for modeling material behavior directly from experimental data [9-13]. While these approaches establish well-structured frameworks that capture complex behaviors effectively, many still rely on predefined model structures or require extensive datasets, limiting their applicability when data acquisition is challenging.

Early studies predominantly relied on parametric identification methods, estimating parameters within predefined constitutive forms such as the Mooney–Rivlin model [14] and the Ogden model [15] for hyperelastic materials. Despite their computational efficiency, these methods depend heavily on prior model assumptions and often fail to capture the full extent of the nonlinear behavior observed in real materials. Moreover, they require extensive experimental data—primarily from uniaxial tension tests—which limits their applicability under complex loading conditions [16-18]. The emergence of full-field measurement techniques, including digital image correlation and digital volume correlation, has enhanced experimental precision and facilitated advanced identification methods such as the virtual fields method and finite element model updating [19-22]. Nevertheless, these approaches remain computationally expensive and highly sensitive to data quality, rendering them less effective when dealing with noisy or incomplete datasets.

Data-driven methodologies have notably expanded the potential for material modeling and constitutive model discovery by mitigating the constraints inherent in conventional, assumption-dependent frameworks. Various methodologies have been explored, including neural network-based techniques [23-25], symbolic regression [26, 27], and data-efficient frameworks such as the Efficient Unsupervised Constitutive Law Identification and Discovery (EUCLID) method [28-33]. Neural networks offer flexibility in capturing complex material behaviors by directly learning stress–strain relationships from experimental data, yet they often require extensive labeled datasets, limiting their applicability in data-scarce environments. The EUCLID framework has significantly advanced the field by inferring constitutive behavior from full-field displacement and reaction force data, reducing dependence on direct stress measurements. Likewise, symbolic regression offers transparent, interpretable models as an alternative to black-box deep learning. Despite these advancements, most existing methods still depend on full-field measurements or substantial datasets, which constrains their applicability in scenarios where only limited experimental data is available.

To overcome the limitations of conventional model identification techniques, this study presents a framework for discovering constitutive models for hyperelastic materials using only scarce measurement data. The framework is built on a physics-informed neural network (PINN) that integrates physical knowledge with data-driven learning [34-37]. Unlike conventional methods that rely on extensive stress–strain or full-field displacement data (**Fig. 1(a)**), the proposed approach successfully discovers constitutive models using only minimal experimental measurements— such as a few pointwise displacements and reaction force

measurements—augmented by embedded physics constraints. By combining PINNs with finite element discretization, we adopt a weak-form formulation with numerical differentiation (**Fig. 1(b)**) rather than strong-form PDEs with automatic differentiation, thereby substantially reducing computational cost without compromising accuracy [38-41]. A two-stage training strategy is employed: an initial solution is obtained via Adam optimization, followed by refinement of both neural network parameters and constitutive model coefficients through L-BFGS and sparse regression.

The key contributions of this study are as follows:

(i) The proposed framework identifies constitutive models using only limited displacement and reaction force data, making it particularly applicable in cases where comprehensive stress–strain measurements are unavailable;

(ii) Unlike traditional parametric methods that require extensive experimental testing, it determines the most appropriate strain energy density function from a single experimental test, thereby enhancing practicality and efficiency;

(iii) The use of a weak-form formulation with finite element discretization, in place of strong-form PDEs, significantly reduces computational burden while maintaining high accuracy.

Overall, the findings indicate that PINN-based approaches can effectively uncover constitutive models under data-limited conditions, offering a viable alternative to conventional methods that depend on extensive datasets and full-field measurements.

The remainder of this paper is organized as follows. **Section 2** introduces the problem formulation, including the governing equations for hyperelastic materials, strain energy density candidates, and the associated objectives and constraints. **Section 3** details the proposed methodology, integrating finite element discretization with the PINN framework and implementing a two-stage training strategy to enhance accuracy and efficiency. **Section 4** presents the results, including data generation, model discovery, and validation of the proposed framework. Finally, **Section 5** summarizes the major outcomes and concludes the study.

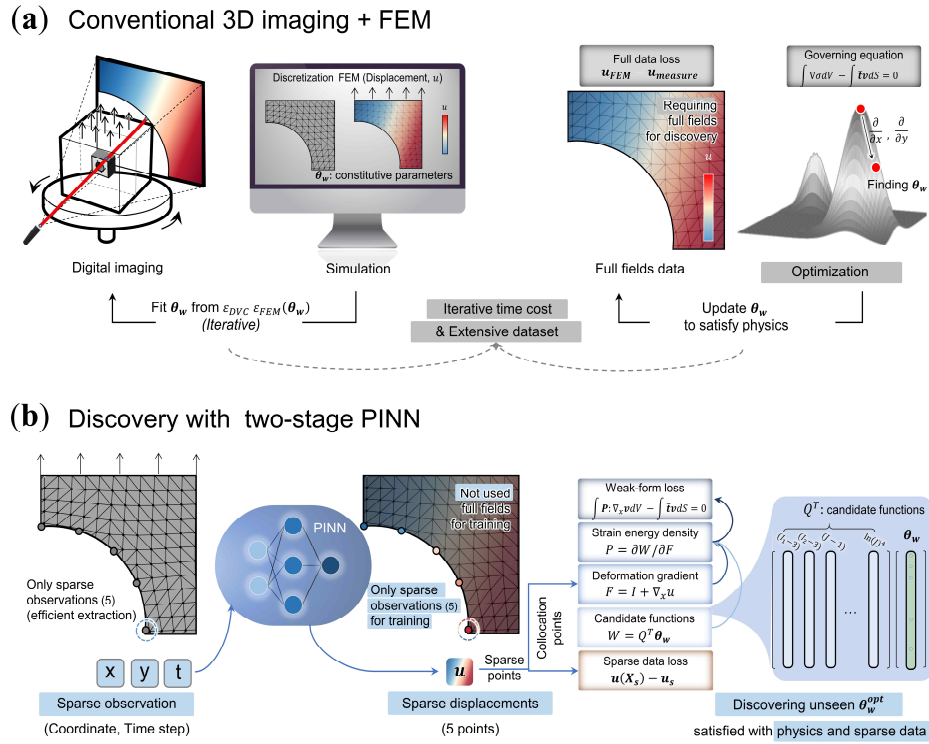

**Fig. 1. Comparison of conventional proposed PINN approaches.** (a) Conventional method: Requires extensive stress–strain or full-field strain data. (b) Proposed PINN method: Leverages limited data with physics constraints

## 2. Problem setting

### 2.1) Governing equations

We begin by considering a static force equilibrium problem governed by the following mechanical equations:

$$\nabla_X \cdot P(X) = 0 \text{ in } X \in \Omega \tag{1a}$$

$$u(X) = \bar{u} \quad on \ X \in \partial\Omega_u \tag{1b}$$

$$P(X) \cdot n(X) = \bar{t} \quad on \ X \in \partial\Omega_t \tag{1c}$$

where $\Omega \in R^3$ is the reference domain, and its boundary $\partial\Omega$ consists of two parts: $\partial\Omega_u$, where Dirichlet boundary conditions are applied, and $\partial\Omega_t$, where Neumann boundary conditions are applied. These boundaries satisfy the conditions $\Omega_u \cup \partial\Omega_t = \partial\Omega$ and $\partial\Omega_u \cap \partial\Omega_t = \emptyset$. Here, $\nabla_X \cdot$ denotes the divergence operator with respect to the reference coordinates, $P$ is the first Piola-Kirchhoff stress tensor, $u$ is the displacement vector, and $n$ is the unit normal vector to the boundary surface in the reference configuration. Assuming negligible body forces, the governing equation in its strong form can be expressed in weak form as follows:

$$\int_\Omega P : \nabla_X v \, dV - \int_{\partial\Omega_t} \bar{t} \cdot v \, dS = 0, \quad \forall \ admissible \ v \tag{2}$$

where $v$ is any admissible displacement field satisfying the Dirichlet boundary conditions.

For hyperelastic materials, the strain energy density $W(F)$ exists as a function of the deformation gradient $F = I + \nabla_X u$, where $\nabla_X$ denotes the gradient operator with respect to the reference coordinates, and $I$ is the second-order identity tensor. The first Piola-Kirchhoff stress tensor $P$ is then determined by the relation $P = \frac{\partial W}{\partial F}$, based on the strain energy density. Moreover, to satisfy the principle of material objectivity (or frame indifference), the strain energy density must be independent of the observer's choice. Consequently, the strain energy density is expressed as a function of the right Cauchy-Green deformation tensor $C = F^T F$.

Assuming an isotropic material, the strain energy density $W$ can be represented in terms of the invariants of $C$ as follows:

$$W = W(I_1, I_2, I_3) \tag{3}$$

where

$$I_1 = tr(C), \quad I_2 = \frac{1}{2}[tr(C)^2 - tr(C^2)], \quad I_3 = \det(C)$$

In this study, we focus on isotropic, nearly compressible hyperelastic materials, where the strain energy density $W$ can be decomposed into an isochoric part $\widehat{W}$ and a volumetric part $\overline{W}$, as follows:

$$W = \widehat{W}(\bar{I}_1, \bar{I}_2) + \overline{W}(J) \tag{4}$$

where $J = \det(F)$, $\bar{I}_1 = I_1/J^{2/3}$, $\bar{I}_2 = I_2/J^{4/3}$. Using this strain energy density, the weak form of the governing equation for hyperelastic materials can be written as:

$$\int_\Omega \frac{\partial W(\bar{I}_1, \bar{I}_2, J)}{\partial F} : \nabla_X v \, dV - \int_{\partial \Omega_t} \bar{t} \cdot v \, dS = 0, \quad \forall \text{ admissible } v \tag{5}$$

Although the current formulation assumes isotropic material behavior, it can be conveniently extended to anisotropic materials by adding an anisotropic strain energy density $W_{aniso}(C, M)$ to Eq. (4), where $M$ denotes structural tensors that define the preferred material directions or symmetries [42-44].

### 2.2) Strain energy density candidates

For hyperelastic materials, discovering the constitutive model is reduced to finding the strain energy density $W$, as the relationship between $P$ and $F$ is given by $P = \frac{\partial W}{\partial F}$. To achieve this, we consider a set $\mathbf{Q}: R^3 \to R^{N_W}$, consisting of several candidates for the strain energy density, where $N_W$ is the total number of candidates in the set. We assume that the strain energy density of the material can be expressed as a linear combination of these candidates. The strain energy density can be written as:

$$W = \boldsymbol{Q}^T(\bar{I}_1, \bar{I}_2, J)\boldsymbol{\theta}_W \tag{6}$$

where $\boldsymbol{\theta}_W \in R^{N_W}$ represents the coefficients of the candidates, which are the unknown parameters to be identified. In principle, any arbitrary strain energy density can be represented by Equation (6) if the candidates set $\boldsymbol{Q}$ is appropriately chosen. In this study, we select the following strain energy density candidates:

$$\boldsymbol{Q} = [\,(\bar{I}_1 - 3)^i (\bar{I}_2 - 3)^{j-i} : j \in \{1, \ldots, N_1\}, i \in \{0, \ldots, j\}\,]$$

$$\oplus \left[(\bar{I}_2^{3/2} - 3^{3/2})^j : j \in \{1, \ldots, N_2\}\right] \oplus \left[(J-1)^{2k} : k \in \{1, \ldots, N_3\}\right]$$

$$\oplus \left[\hbar\,(J)^{2k} : k \in \{1, \ldots, N_4\}\right] \oplus [J - 1 - \hbar\,(J)] \tag{7}$$

Here, $\oplus$ represents the concatenation of sets. The first and second candidate sets are used for the isochoric part of the strain energy density $\widehat{W}(\bar{I}_1, \bar{I}_2)$, while the remaining sets are used for the volumetric part $\overline{W}(J)$. The first set describes the generalized Mooney-Rivlin model, while the second set is inspired by the Hartmann-Neff model. For the volumetric strain energy density, we selected the most commonly used polynomial form (third set) and logarithmic forms (fourth and fifth sets).

To provide a clearer understanding of the proposed formulation, we first consider a simplified case where the number of candidate functions is limited to $N_1 = 1, N_2 = 1, N_3 = 1, N_4 = 1$. Under these conditions, the candidate set $\boldsymbol{Q}$ is given by:

$$\boldsymbol{Q} = [(\bar{I}_1 - 3), (\bar{I}_2 - 3), (\bar{I}_1 - 3)^2, (\bar{I}_2^{3/2} - 3^{3/2}), (J-1)^2, \hbar\,(J)^2, (J - 1 - \hbar\,(J))]$$

Consequently, the strain energy density function can be expressed as a linear combination of these candidates:

$$W = \theta_1(\bar{I}_1 - 3) + \theta_2(\bar{I}_2 - 3) + \theta_3(\bar{I}_1 - 3)^2 + \theta_4\left(\bar{I}_2^{\frac{3}{2}} - 3^{\frac{3}{2}}\right) + \theta_5(J-1)^2 + \theta_6 h(J)^2 + \theta_7(J - 1 - h(J))$$

This formulation illustrates how the strain energy density function is constructed from a predefined set of candidate functions. Extending this approach, we set $N_1 = N_2 = N_3 = N_4 = 7$ in this study, resulting in total $N_W = 57$ candidates (see **Supplementary Note 1** for details). Building upon this formulation, the first Piola-Kirchhoff stress can be written in index notation as:

$$P_{ij} = \frac{\partial W(\bar{I}_1, \bar{I}_2, J)}{\partial F_{ij}} = \frac{\partial \boldsymbol{Q}^T(\bar{I}_1, \bar{I}_2, J)}{\partial F_{ij}} \boldsymbol{\theta}_W \tag{8}$$

where the derivative term $\frac{\partial \boldsymbol{Q}^T(\bar{I}_1, \bar{I}_2, J)}{\partial \boldsymbol{F}}$ can be analytically derived by applying the chain rule, given the deformation gradient $\boldsymbol{F}$, based on the following differentiation relations.

$$\frac{\partial \bar{I}_1}{\partial \boldsymbol{F}} = \frac{2}{J^{\frac{2}{3}}} \boldsymbol{F} - \frac{2}{3} \bar{I}_1 \boldsymbol{F}^{-T} \tag{9a}$$

$$\frac{\partial \bar{I}_2}{\partial \boldsymbol{F}} = \frac{2}{J^{\frac{2}{3}}} \bar{I}_1 \boldsymbol{F} - \frac{2}{J^{\frac{4}{3}}} \boldsymbol{F} \cdot \boldsymbol{F}^T \cdot \boldsymbol{F} - \frac{4}{3} \bar{I}_2 \boldsymbol{F}^{-T} \tag{9b}$$

$$\frac{\partial J}{\partial \boldsymbol{F}} = J \boldsymbol{F}^{-T} \tag{9c}$$

Furthermore, to ensure physical consistency, the strain energy density must satisfy the following well-known, physically reasonable constraints [45-47]:

(i) Non-negativity of strain energy density.
$$W = \widehat{W}(\bar{I}_1, \bar{I}_2) + \overline{W}(J) \geq 0 \quad \forall \boldsymbol{F}$$

(ii) Normalization of stress and strain energy density.
$$W = \widehat{W}(\bar{I}_1, \bar{I}_2) + \overline{W}(J) = 0 \text{ and } \boldsymbol{P} = \boldsymbol{0} \quad \text{at } \boldsymbol{F} = \boldsymbol{I}$$

(iii) Growth condition.
$$W = \widehat{W}(\bar{I}_1, \bar{I}_2) + \overline{W}(J) \to \infty \quad \text{for } J \to 0 \text{ or } J \to \infty$$

The strain energy density candidates in (7) automatically satisfy the first two conditions because all candidate sets are chosen from the strain energy density function. To satisfy the third condition, we assume that all coefficients $\boldsymbol{\theta_W}$ are non-negative.

## 2.3) Problem definition

The objective of this study is to discover the constitutive model of hyperelastic materials using extremely scarce measurement data obtained from quasi-static material testing experiments. Specifically, this process reduces to selecting the most suitable candidates from a set of predefined strain energy density candidates and identifying their corresponding coefficients. By substituting equation (8) into equation (5), the coefficients $\boldsymbol{\theta_W}$ can be determined to satisfy the following force equilibrium condition:

$$\int_\Omega \left( \frac{\partial \boldsymbol{Q}^T(\bar{I}_1, \bar{I}_2, J)}{\partial F_{ij}} \boldsymbol{\theta_W} \right) v_{i,j} \, dV - \int_{\partial \Omega_t} \bar{t}_i \cdot v_i \, dS = 0, \qquad \forall \text{ admissible } v \qquad (10)$$

Assuming that the material is homogeneous, $\boldsymbol{\theta_W}$ is treated as a constant independent of coordinates, so the equation simplifies as:

$$\left( \int_\Omega \frac{\partial \boldsymbol{Q}^T(\bar{I}_1, \bar{I}_2, J)}{\partial F_{ij}} v_{i,j} \, dV \right) \boldsymbol{\theta_W} - \int_{\partial \Omega_t} \bar{t}_i \cdot v_i \, dS = 0, \qquad \forall \text{ admissible } v \qquad (11)$$

In displacement-controlled material testing, if reaction force data and full-field displacement data are available, the deformation gradient field $\boldsymbol{F}$ can be directly computed by differentiating the displacement field. Using the deformation gradient field in the first term of equation (11) and reaction force data in the second term, $\boldsymbol{\theta_W}$ can be determined by solving the following optimization problem:

$$\boldsymbol{\theta_W^*} = \underset{\boldsymbol{\theta_W}}{\arg\min} \left\| \left( \int_\Omega \frac{\partial \boldsymbol{Q}^T(\bar{I}_1, \bar{I}_2, J)}{\partial F_{ij}} v_{i,j} \, dV \right) \boldsymbol{\theta_W} - \left( \int_{\partial \Omega_t} \bar{t}_i \cdot v_i \, dS \right) \right\|^2 \qquad (13)$$

Previous studies, including the EUCLID method, have shown that it is possible to discover constitutive models using full-field displacement data obtained through techniques such as Digital Image Correlation (DIC), even without direct stress measurements. While effective, these approaches typically rely on complex and resource-intensive experimental setups, which may not be practical in many scenarios. To overcome this limitation, the present study focuses on discovering constitutive models using more accessible data, specifically reaction forces and displacements measured at a limited number of discrete points on the specimen. The key challenge, therefore, is to reconstruct the displacement field $u$ such that it satisfies the force equilibrium equation while remaining consistent with the available experimental data. This leads to the following multi-objective optimization problem:

$$u^*, \boldsymbol{\theta}_W^* = \underset{u, \theta_W}{arg\,min} \left\| \left( \int_\Omega \frac{\partial \boldsymbol{Q}^T(\bar{I}_1, \bar{I}_2, J)}{\partial F_{ij}} v_{i,j}\, dV \right) \boldsymbol{\theta}_W - \left( \int_{\partial \Omega_t} \bar{t}_i \cdot v_i\, dS \right) \right\|^2 + \lambda \sum_{m=1}^{N_m} \|u(X_m) - u_m\|^2 \quad (14)$$

where $X_m$ represents the $m$-th measurement point in the reference domain, and $u_m$ is the displacement value measured at that point. $N_m$ denotes the number of measurement points on the specimen, and $\lambda$ is a weight that adjusts the balance between the objectives. Additionally, data from multiple loading steps obtained during the experiment can be used, so the optimization problem can be extended to incorporate data from multiple loading steps:

$$u^*, \boldsymbol{\theta}_W^* = \underset{u, \theta_W}{arg\,min} \sum_{t=1}^{N_T} \left( \left\| \left( \int_\Omega \frac{\partial \boldsymbol{Q}^T(\bar{I}_1, \bar{I}_2, J)}{\partial F_{ij}} v_{i,j}\, dV \right)^t \boldsymbol{\theta}_W - \left( \int_{\partial \Omega_t} \bar{t}_i \cdot v_i\, dS \right)^t \right\|^2 + \lambda \sum_{m=1}^{N_m} \|u^t(X_m) - u_m^t\|^2 \right) \quad (15)$$

where the superscript $t$ indicates that the respective terms are evaluated at loading step $t$, and $N_T$ denotes the total number of loading steps in the experiment.

Consequently, the problem addressed in this study is to solve the optimization problem in equation (15) to simultaneously reconstruct the full-field displacement $u$ from scarce measurement data and identify the constitutive parameters $\boldsymbol{\theta}_W$. Notably, the available data consist solely of reaction forces and displacements at a few discrete locations, without any direct strain or stress measurements. Furthermore, the exact governing partial differential equations (PDEs) remain unknown because the constitutive model itself is undefined. Thus, the challenge is to formulate and solve an inverse problem that, using only scarce and localized data, predicts the full-field displacement across the domain while determining the constitutive model that governs the PDEs. This problem is inherently ill-posed due to the scarcity of data, requiring careful methodological design to ensure its solvability. To address this challenge, we introduce the PINN framework, which integrates physical constraints with data-driven learning to infer both the displacement field and the underlying constitutive model. The details of this approach will be discussed in **Section 3**.

## 3. Methodology

In **Section 3**, we detail the methodology used to address the problem outlined in **Section 2.3**. The domain is discretized using the finite element method (FEM) with triangular meshes, and PINN predicts nodal displacements to reconstruct the full-field displacement. During PINN training, both the strain energy density coefficients and the neural network parameters are treated as optimization variables, determined by minimizing the loss function in Equation (15). A two-stage training strategy, described later in this section, is employed to enhance performance. For clarity, the neural network parameters are denoted as $\boldsymbol{\theta}_{NN}$, and the strain energy density coefficients, which represent the constitutive parameters to be identified, are denoted as $\boldsymbol{\theta}_W$. The notation employed herein follows the conventions established in [28,29].

### 3.1) Finite element discretization

To solve the optimization problem defined in Equation (15), the reference domain $\Omega$ of the material testing specimen is discretized using finite elements. This process converts the integral form of the force equilibrium equation into algebraic equations, allowing the optimization problem to be reformulated. The domain $\Omega$ is divided into $N_n$ nodes, and the degrees of freedom (DOF) for the three-dimensional problem total $3N_n$. The set of all DOFs is defined as $D = \{(a, i): a = 1, \ldots, N_n; i = 1,2,3\}$, where $a$ denotes node numbering and $i$ indicates spatial dimensions (i.e., $x_1, x_2, x_3$). As shown in **Fig. 2**, the set $D$ is further partitioned into $D^{fix}$, representing DOFs with Dirichlet boundary conditions, and the remaining DOFs $D^{free} = D \setminus D^{fix}$.

Under displacement-controlled conditions, reaction forces are applied only to $D^{fix}$, while $D^{free}$ experiences zero net force. Using finite element shape functions, the

displacement $u(X)$ is approximated as:

$$u(X) = \sum_{a=1}^{N_n} N^a(X) u^a \tag{16}$$

where $N^a : \Omega \to R$ is the shape function for node $a$ and $u^a$ denotes the displacement value at node. Similarly, the admissible displacement field $v(X)$ is discretized as:

$$v(X) = \sum_{a=1}^{N_n} N^a(X) v^a \quad wth \; v_i^a = 0 \, f \; (a,i) \in D^{fix} \tag{17}$$

The deformation gradient is expressed as:

$$F(X) = I + \sum_{a=1}^{N_n} u^a \otimes \nabla_X N^a(X) \tag{18}$$

Using the equation (16) and (17), the force equilibrium equation (11) can be written as:

$$\sum_{a=1}^{N_n} v^a \left[ \left( \int_\Omega \frac{\partial Q^T(\bar{I}_1, \bar{I}_2, J)}{\partial F_{ij}} (\nabla_X N^a)_j \, dV \right) \theta_W - \int_{\partial \Omega_t} \bar{t}_i \cdot N^a \, dS \right] = 0 \tag{19}$$

For $D^{free}$ where net force is zero, this reduces to:

$$\left( \int_\Omega \frac{\partial Q^T(\bar{I}_1, \bar{I}_2, J)}{\partial F_{ij}} (\nabla_X N^a)_j \, dV \right) \theta_W = 0, \quad \forall \; (a,i) \in D^{free} \tag{20}$$

This equation is linear in $\theta_W$ and can be assembled into a matrix form:

$$A^{free} \theta_W = 0 \quad where \quad A^{free} \in R^{|D^{free}| \times N_W} \tag{21}$$

For $D^{fix}$, the internal force equals the reaction force:

$$\left( \int_\Omega \frac{\partial Q^T(\bar{I}_1, \bar{I}_2, J)}{\partial F_{ij}} (\nabla_X N^a)_j \, dV \right) \theta_W = r_i^a, \quad \forall \; (a,i) \in D^{fix} \tag{22}$$

where $r_i^a$ is the reaction force corresponding to the $i$-th degree of freedom of the $a$-th node. In the displacement-controlled material testing experiments, it is difficult to know the reaction force at every point on the surface $\partial\Omega_u$. Instead, we typically measure the resultant reaction force acting on each surface corresponding to a loaded side of the specimen in one direction. These degrees of freedom form a subset of $D^{fix}$. Let $N_\alpha$ be the number of such subsets, and let $D^{fix,\alpha}$ represent these subsets, where $\alpha = \{1, ..., N_\alpha\}$, with $D^{fix,\alpha} \subseteq D^{fix}$ and $\cup_{\alpha=1}^{N_\alpha} D^{fix,\alpha} = D^{fix}$, and $D^{fix,\alpha} \cap D^{fix,\beta} = \emptyset \; for \; \alpha \neq \beta$. For each subset $D^{fix,\alpha}$, the force equilibrium condition is expressed as:

$$\sum_{(a,i)\in D^{fix,\alpha}} \left( \int_\Omega \frac{\partial \boldsymbol{Q}^T(\bar{I}_1, \bar{I}_2, J)}{\partial F_{ij}} (\nabla_X N^a)_j \, dV \right) \boldsymbol{\theta}_W = R^\alpha, \quad \alpha = 1, ..., N_\alpha \tag{23}$$

where $R^\alpha$ is the resultant reaction force in the direction of displacement control for the surface corresponding to the $a$-th subset $D^{fix,\alpha}$, which can be obtained through measurement in the experiment.

This can also be written in matrix form:

$$\boldsymbol{A}^{fix} \boldsymbol{\theta}_W = \boldsymbol{b}^{fix} \quad where \quad \boldsymbol{A}^{fix} \in R^{N_\alpha \times N_c}, \boldsymbol{b}^{fix} \in R^{N_\alpha} \tag{24}$$

The matrices $\boldsymbol{A}^{free}$ and $\boldsymbol{A}^{fix}$ can be computed numerically using Gaussian quadrature for the integrals, and the vector $\boldsymbol{b}^{fix}$ is composed of the resultant reaction forces.

By discretizing the force equilibrium equations and converting them into algebraic equations, the optimization problem equation (14) can be reformulated as:

$$\boldsymbol{u}^*, \boldsymbol{\theta}^* = argmin \; \|\boldsymbol{A}^{free} \boldsymbol{\theta}_W\|^2 + \|\boldsymbol{A}^{fix} \boldsymbol{\theta}_W - \boldsymbol{b}^{fix}\|^2 + \lambda \sum_{m=1}^{N_m} \|\boldsymbol{u}(X_m) - \boldsymbol{u}_m\|^2 \tag{25}$$

When multiple data points from different load steps are available, the optimization problem can be similarly extended as follows:

$$\boldsymbol{u}^*, \boldsymbol{\theta}_W^* = \underset{u,\theta_W}{argmin} \sum_{t=1}^{N_T} \left( \|\boldsymbol{A}^{free,t} \boldsymbol{\theta}_W\|^2 + \|\boldsymbol{A}^{fix,t} \boldsymbol{\theta}_W - \boldsymbol{b}^{fix,t}\|^2 + \lambda \sum_{m=1}^{N_m} \|\boldsymbol{u}^t(X_m) - \boldsymbol{u}_m^t\|^2 \right) \tag{26}$$

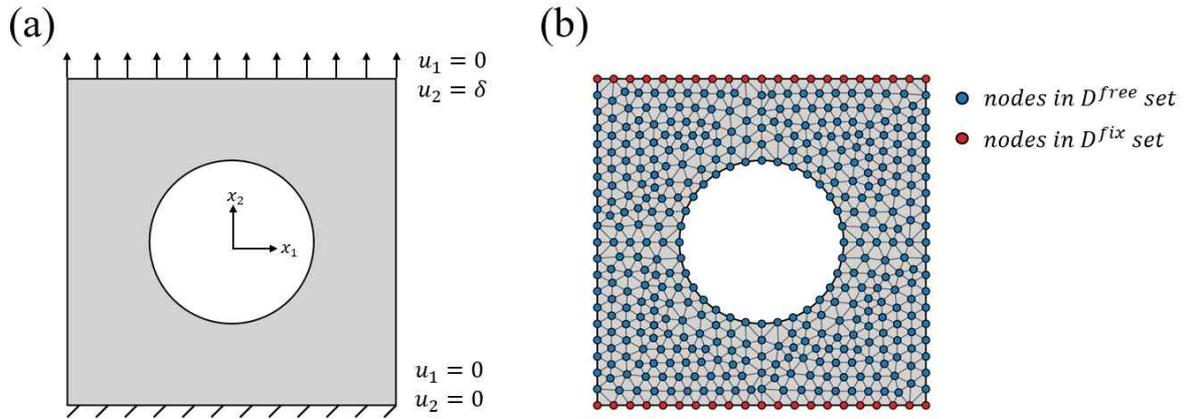

**Fig. 2. Example of domain discretization by finite element method.** (a) Schematic of the specimen and boundary conditions: displacement-controlled loading. (b) Finite element discretization, highlighting $D^{free}$ (blue) and $D^{fix}$ (red) nodes.

By employing discretization, we can compute gradients at a lower computational cost compared to other differentiation methods, such as numerical or automatic differentiation. This approach is motivated by the need to perform differentiation for a large number of strain energy density candidates. Using numerical or automatic differentiation would result in prohibitively high computational costs due to the extensive number of operations involved. Therefore, we opted for finite element discretization, which enables gradient computation using shape functions, significantly reducing the computational cost.

## 3.2) Physics-Informed Neural Networks (PINN)

In this study, we employ PINN to address the inverse problem of reconstructing the displacement field of the material testing specimen using the given measurement data. The neural network takes as input the reference coordinates $X$ of the nodes used for discretizing the domain and the loading step $t = \{1, ..., N_T\}$, and it predicts the displacement at each position for the corresponding loading step. For a neural network with $N_{layer}$ hidden layers, the following computations are performed:

$$\boldsymbol{z_0} = [X^a, t]^T \tag{27a}$$

$$\boldsymbol{z_i} = \phi(\boldsymbol{W_i}\boldsymbol{z_{i-1}} + \boldsymbol{b_i}) \quad with \quad i \in \{1, ..., N_{layer}\} \tag{27b}$$

$$\boldsymbol{u}(X^a, t) = \boldsymbol{z_{N+1}} = \boldsymbol{W_{N+1}}\boldsymbol{z_N} + \boldsymbol{b_{N+1}} \tag{27c}$$

Here, $z_i$ represents the output of the $i$-th hidden layer, while $\boldsymbol{W_i}$, $\boldsymbol{b_i}$ are the weights and biases of the $i$-th hidden layer, respectively. These are the neural network parameters, denoted as $\boldsymbol{\theta_{NN}}$, which are updated during training. The activation function is represented by $\phi$. In this study, the PINN model was implemented using PyTorch.

## 3.3) Two-stage training strategy

In this study, we have two types of parameters to update through training: the network parameters $\boldsymbol{\theta_{NN}}$ and the constitutive parameters $\boldsymbol{\theta_W}$. The training process incorporates losses: physics loss ($L_{physics}$), data loss ($L_{data}$), and regularization loss ($L_{sparse}$) which are expressed as follow:

$$L_{physics} = \sum_{t=1}^{N_T} \left( \|A^{free,t}\boldsymbol{\theta_W}\|^2 + \|A^{fix,t}\boldsymbol{\theta_W} - b^{fix,t}\|^2 \right) \tag{28a}$$

$$L_{data} = \lambda \sum_{t=1}^{N_T} \left( \sum_{m=1}^{N_m} \|\boldsymbol{u}^t(\boldsymbol{X}_m) - \boldsymbol{u}_m^t\|^2 \right) \tag{28b}$$

$$L_{sparse} = \lambda_p \|\boldsymbol{\theta}_W\|_p^p = \lambda_p \sum_{i=1}^{N_W} |\theta_{W,i}|^p \tag{28c}$$

where $\theta_{W,i}$ is the $i$-th component of $\boldsymbol{\theta}_W$, $\lambda_p$ is the weight for the regularization loss, and $p$ is the regularization exponent. The physics loss $L_{physics}$ ensures that the predicted displacement field satisfies the force equilibrium equations, while the data loss $L_{data}$ aligns the displacement predictions with measured values at specific points. Additionally, the regularization loss $L_{sparse}$ enforces sparsity in the constitutive parameters, promoting a parsimonious model by encouraging most of the components of $\boldsymbol{\theta}_W$ to be zero or near zero. By incorporating regularization loss, this approach encourages sparsity, helping to identify a small subset of the most relevant strain energy density candidates that best describe the material behavior among the many initially considered. Notably, the sum of $L_{physics}$ and $L_{data}$ constitutes the objective function for the optimization problem in equation (26).

The training process is divided into two stages (as shown in **Fig. 3**). In the first stage, both the network parameters $\boldsymbol{\theta}_{NN}$ and the constitutive parameters $\boldsymbol{\theta}_W$ are optimized simultaneously using the Adam optimizer. The loss function for this stage is the sum of $L_{physics}$ and $L_{data}$. Unlike conventional PINN, which rely on explicitly known governing partial differential equations (PDEs), this approach infers the PDE indirectly through the data, as the constitutive parameters are initially unknown. This simultaneous optimization allows the model to learn both the neural network parameters and the constitutive parameters. The optimization problem for this stage is expressed as:

$$\boldsymbol{\theta}_{NN}^*, \boldsymbol{\theta}_W^* = \underset{\boldsymbol{\theta}_{NN}, \boldsymbol{\theta}_W}{\arg\min} \ L_{physics} + L_{data} \tag{29}$$

In the second stage, training alternates between updating the network parameters and the constitutive parameters. The network parameters $\theta_{NN}$ are updated using the L-BFGS optimizer to minimize $L_{physics}$ and $L_{data}$. For the constitutive parameters $\theta_W$, sparse regression is employed, integrating the physics loss $L_{sparse}$ with regularization loss $L_{sparse}$. This results in a regularized least-squares optimization problem with respect to $\theta_W$, designed to promote sparsity in the solution. Sparse regression is performed using the fixed-point iteration method [28, 29], enabling the identification of the most relevant constitutive parameters. The optimization problems for this stage are expressed as:

$$\theta_{NN}^* = \operatorname*{argmin}_{\theta_{NN}} \quad L_{physics} + L_{data} \tag{30a}$$

$$\theta_{W}^* = \operatorname*{argmin}_{\theta_{W}} \quad L_{physics} + L_{sparse} \tag{30b}$$

This two-stage training approach effectively reconstructs the displacement field and discover the constitutive model (i.e., the strain energy density) using limited observed data. The first stage employs the Adam optimizer to approximate a solution near the desired values, while the second stage refines the solution using L-BFGS and sparse regression. This ensures that the force equilibrium is satisfied and results in a parsimonious constitutive model.

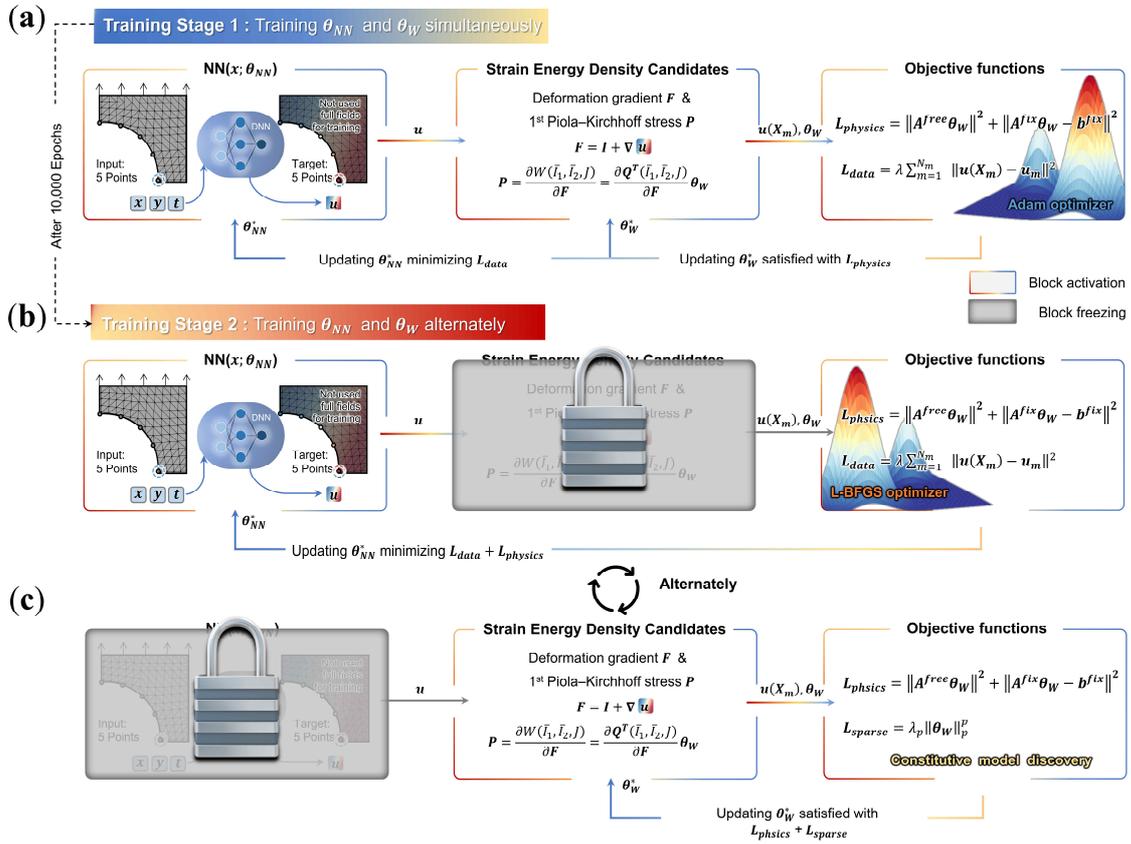

**Fig. 3. Schematic of the two-stage training strategy.** (a) Stage 1: Simultaneous optimization of neural network parameters ($\theta_{NN}$) and constitutive model parameters ($\theta_W$) using the Adam optimizer. (b, c) Stage 2: Alternating optimization of $\theta_{NN}$ updated with the L-BFGS optimizer and $\theta_W$ refined through sparse regression to enforce parsimony.

## 4. Results and discussion

### 4.1) Data generation

To validate the proposed method, virtual experimental data generated through Finite Element Analysis (FEA) were used in place of actual experimental data from material testing. The material specimen analyzed in this study is a hyperelastic plate with a central hole, as illustrated in **Fig. 4**. Due to the symmetry of the geometry, only one-quarter of the plate is modeled for analysis. The specimen undergoes a triaxial tension-compression test, with displacement-controlled loading applied incrementally over 10 loading steps: tension in two in-plane directions and compression in the thickness direction. Since the displacement is uniform in the thickness direction, the specimen can be considered to be under plane strain conditions at each loading step.

The reaction forces on the right and top surfaces are extracted from the force-displacement curve, while displacement data at the five red points shown in **Fig. 4(a)** are obtained. These data are incorporated in the $L_{physics}$ and $L_{data}$, respectively. This combination of geometry and loading was specifically chosen, in contrast to traditional biaxial tension or torsion tests, because the aim is to solve an inverse problem using scarce measurement data, without stress or strain data, and based on a single test. The chosen geometry and loading configuration are considered highly suitable for addressing this challenging problem of reconstructing the displacement field and discovering the constitutive model.

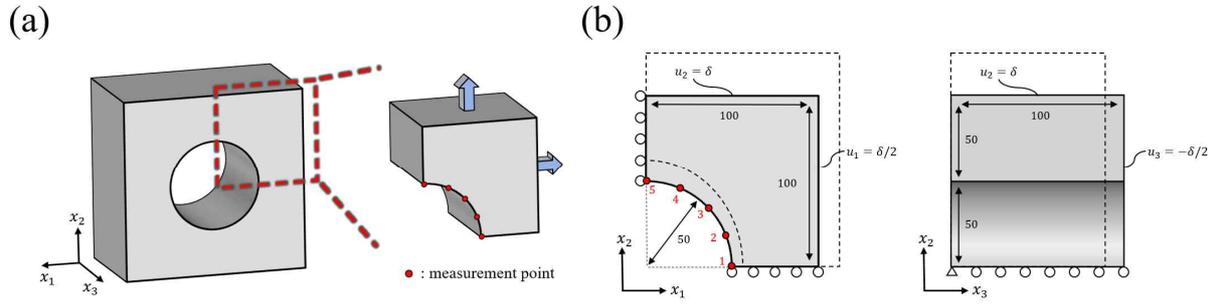

**Fig. 4. Virtual experiment specimen: geometry, boundary conditions, and measurement points.** (a) Geometry of the hyperelastic plate with symmetry considerations. (b) Displacement-controlled loading with loading parameter $\delta$ and measurement point locations.

The virtual experiments via FEA were carried out using the following benchmark hyperelastic models, which are widely used in the literature. Data generation was performed for each of these models to facilitate the validation of the proposed method [28].

- NH2: Neo-Hookean solid with quadratic volumetric strain energy

$$W = 0.5(\bar{I}_1 - 3) + 1.5(J - 1)^2$$

- NH4: Neo-Hookean solid with biquadratic volumetric strain energy

$$W = 0.5(\bar{I}_1 - 3) + 1.5(J - 1)^4$$

- IH: Isihara solid with quadratic deviatoric strain energy

$$W = 0.5\,(\bar{I}_1 - 3) + 1.0\,(\bar{I}_2 - 3) + 1.0\,(\bar{I}_1 - 3)^2 + 1.5(J - 1)^2$$

- HW: Haines–Wilson solid with cubic deviatoric strain energy

$$W = 0.5(\bar{I}_1 - 3) + 1.0(\bar{I}_2 - 3) + 0.7(\bar{I}_1 - 3)(\bar{I}_2 - 3) + 0.2(\bar{I}_1 - 3)^3 + 1.5(J - 1)^2$$

- MR2: Mooney-Rivlin solid with quadratic logarithmic volumetric strain energy

$$W = 0.5(\bar{I}_1 - 3) + 1.0(\bar{I}_2 - 3) + 1.5\,\hbar\,(J)^2$$

- MR4: Mooney-Rivlin solid with biquadratic logarithmic volumetric strain energy

$$W = 0.5\,(\bar{I}_1 - 3) + 1.0\,(\bar{I}_2 - 3) + 1.5\,\hbar\,(J)^4$$

- YH: Yeoh solid with logarithmic volumetric strain energy

$$W = 0.5\,(\bar{I}_1 - 3) + 0.7\,(\bar{I}_1 - 3)^2 + 0.3\,(\bar{I}_1 - 3)^3 + 1.5\,ℏ\,(J)^2$$

- HN: Hartmann-Neff Solid with volumetric strain energy

$$W = 0.5\,(\bar{I}_1 - 3) + 1.0\,(\bar{I}_2^{3/2} - 3^{3/2})^2 + 1.5(J - 1 - ℏ\,(J))$$

While the data obtained through deterministic FEM simulations are noise-free, real experimental data are inevitably affected by noise. To evaluate the robustness of the proposed framework to noisy measurement data, Gaussian noise with a mean of zero and standard deviations of 1%, 3%, and 5% of the FEM data values was added to the FEM-generated data from all loading steps (as shown in **Fig. 4**). The framework's ability to discover the constitutive model was then tested using these noisy measurement data at varying noise levels.

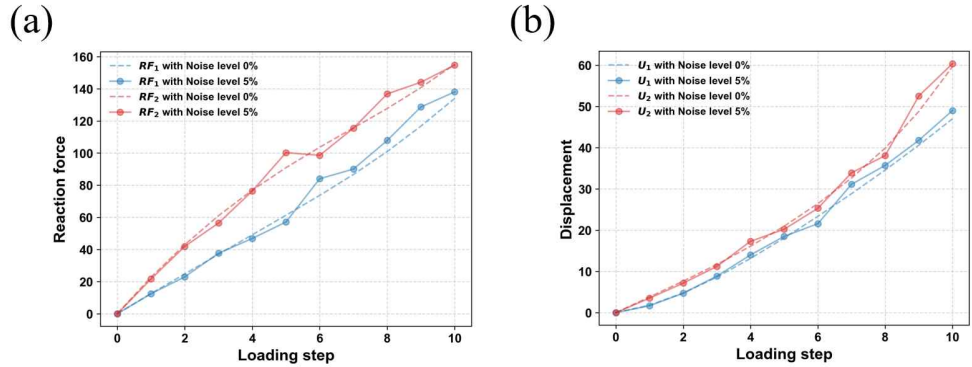

**Fig. 4**. **Measured dataset at 0% and 5% Gaussian noise levels across loading steps** (a) Resultant reaction force data (b) displacement data for measurement point 3.

### 4.2) Discovery results

All subsequent results are based on the parameters, including algorithm hyperparameters, listed in **Table 1**, which were determined through a careful testing phase. The

PINN architecture utilized in this study consists of fully connected hidden layers, each containing 64 nodes, with Sigmoid Linear Unit (SiLU) activation functions applied after each layer. The network takes the node coordinates from the discretized domain and the loading step as input, and outputs the displacement at each node.

**Table 1.** Hyperparameters for the data generation and the proposed method

| Parameter | Notation | Value |
|---|---|---|
| Number of nodes in the mesh for domain discretization | $N_n$ | 2185 |
| Number of surfaces for reaction force acquisition | $N_\alpha$ | 4 |
| Number of measurement points for displacement data | $N_m$ | 5 |
| Loading parameter | $\delta$ | 100 |
| Number of loading steps | $N_T$ | 10 |
| Number of strain energy density candidates | $N_W$ | 57 |
| Weight of the data loss | $\lambda$ | 0.5 |
| Weight of the regularization loss | $\lambda_p$ | 1e-3 |
| Exponent of the regularization loss | $p$ | 0.25 |
| Learning epochs in training Stage 1 (Adam optimizer) | - | 20000 |
| Maximum epochs in training Stage 2 (L-BFGS optimizer) | - | 2000 |
| Number of alternate training cycles in training Stage 2 | - | 10 |
| Number of hidden layers in the neural network | $N_{layer}$ | 6 |
| Number of nodes in each hidden layer of the neural network | - | 64 |
| Activation function used in the neural network | $\phi$ | SiLU |

The discovery results for the strain energy density of the eight benchmark models using the proposed method are summarized in **Table 2**. When clean FEM data with 0% noise is used, the framework accurately select the correct strain energy density candidates and coefficients for all eight models, with only minor deviations in the coefficients that do not significantly affect the material behavior. At 1% noise, the framework continues to perform effectively, correctly discovering the strain energy density candidates and coefficients that closely match the true function. At 3% noise, the framework remains robust for most models, successfully selecting the correct candidates and coefficients except for the HN model. In the case of the HN model, similar candidates, such as $\left(\bar{I}_2^{3/2} - 3^{3/2}\right)^2$, $(\bar{I}_2 - 3)^2$, and $(\bar{I}_2 - 3)^3$, appear to cause confusion due to the impact of noise. When noise increases to 5%, the framework's performance begins to degrade; however, models such as MR2, MR4, and YH are still correctly discovered. Considering that 3% noise is relatively high in typical material testing experiments, these results highlight the effectiveness of the proposed framework in discovering constitutive models from real experimental data, even under conditions of data scarcity and noise.

**Table 2.** Strain energy density discovery results for benchmark models with different noise level

| Benchmarks | | | Strain energy density (W) |
|---|---|---|---|
| **NH2** | True | | $0.500\,(\bar{I}_1 - 3) + 1.500\,(J - 1)^2$ |
| | Noise level | 0% | $0.49722(\bar{I}_1 - 3) + 1.49448(J - 1)^2$ |
| | | 1% | $0.49805(\bar{I}_1 - 3) + 1.49138(J - 1)^2$ |
| | | 3% | $0.49561(\bar{I}_1 - 3) + 1.50906(J - 1)^2$ |
| | | 5% | $0.43805(\bar{I}_1 - 3) + 0.04785(\bar{I}_2 - 3) + 0.00372(\bar{I}_1 - 3)^2(\bar{I}_2 - 3) + 1.81354\ln(J)^2$ |
| **NH4** | True | | $0.500\,(\bar{I}_1 - 3) + 1.500\,(J - 1)^4$ |
| | Noise level | 0% | $0.49693\,(\bar{I}_1 - 3) + 1.51208(J - 1)^4$ |
| | | 1% | $0.49952(\bar{I}_1 - 3) + 1.48065(J - 1)^4$ |
| | | 3% | $0.49593(\bar{I}_1 - 3) + 1.53003(J - 1)^4$ |
| | | 5% | $0.47715(\bar{I}_1 - 3) + 1.91826(J - 1)^6 + 2.34429\,\ln(J)^4$ |
| **IH** | True | | $0.500\,(\bar{I}_1 - 3) + 1.000\,(\bar{I}_2 - 3) + 1.000\,(\bar{I}_1 - 3)^2 + 1.500\,(J - 1)^2$ |
| | Noise level | 0% | $0.46284(\bar{I}_1 - 3) + 1.02096(\bar{I}_2 - 3) + 1.00504(\bar{I}_1 - 3)^2 + 1.48155(J - 1)^2$ |
| | | 1% | $0.46517(\bar{I}_1 - 3) + 0.99961(\bar{I}_2 - 3) + 1.00376(\bar{I}_1 - 3)^2 + 1.49895(J - 1)^2$ |
| | | 3% | $0.43036(\bar{I}_1 - 3) + 0.99335(\bar{I}_2 - 3) + 1.02514(\bar{I}_1 - 3)^2 + 1.48625(J - 1)^2$ |
| | | 5% | $0.14496(\bar{I}_1 - 3) + 1.29874(\bar{I}_2 - 3) + 1.03447(\bar{I}_1 - 3)^2 + 1.75937\,\ln(J)^4$ |
| **HW** | True | | $0.500\,(\bar{I}_1 - 3) + 1.000\,(\bar{I}_2 - 3) + 0.700\,(\bar{I}_1 - 3)(\bar{I}_2 - 3) + 0.200\,(\bar{I}_1 - 3)^3 + 1.500\,(J - 1)^2$ |
| | Noise level | 0% | $0.482(\bar{I}_1 - 3) + 1.021(\bar{I}_2 - 3) + 0.692(\bar{I}_1 - 3)(\bar{I}_2 - 3) + 0.201(\bar{I}_1 - 3)^3 + 1.483(J - 1)^2$ |
| | | 1% | $0.509(\bar{I}_1 - 3) + 1.975(\bar{I}_2 - 3) + 0.718(\bar{I}_1 - 3)(\bar{I}_2 - 3) + 0.192(\bar{I}_1 - 3)^3 + 1.496(J - 1)^2$ |
| | | 3% | $0.487(\bar{I}_1 - 3) + 0.998(\bar{I}_2 - 3) + 0.718(\bar{I}_1 - 3)(\bar{I}_2 - 3) + 0.181(\bar{I}_1 - 3)^3 + 1.432(J - 1)^2$ |
| | | 5% | $1.751(\bar{I}_2 - 3) + 0.457(\bar{I}_1 - 3)^2 + 0.264(\bar{I}_1 - 3)^2 + 0.014\left(\bar{I}_2^{3/2} - 3^{3/2}\right)^2 + 1.535(J - 1)^2$ |
| **MR2** | True | | $0.500\,(\bar{I}_1 - 3) + 1.000\,(\bar{I}_2 - 3) + 1.500\ln(J)^2$ |
| | Noise level | 0% | $0.495(\bar{I}_1 - 3) + 1.001(\bar{I}_2 - 3) + 1.487\ln(J)^2$ |
| | | 1% | $0.497(\bar{I}_1 - 3) + 1.003(\bar{I}_2 - 3) + 1.494\ln(J)^2$ |
| | | 3% | $0.492(\bar{I}_1 - 3) + 1.017(\bar{I}_2 - 3) + 1.526\ln(J)^2$ |
| | | 5% | $0.486(\bar{I}_1 - 3) + 1.986(\bar{I}_2 - 3) + 1.491\ln(J)^2$ |
| **MR4** | True | | $0.500\,(\bar{I}_1 - 3) + 1.000\,(\bar{I}_2 - 3) + 1.500\ln(J)^4$ |
| | Noise level | 0% | $0.485(\bar{I}_1 - 3) + 1.027(\bar{I}_2 - 3) + 1.08592\ln(J)^4$ |
| | | 1% | $0.479(\bar{I}_1 - 3) + 1.035(\bar{I}_2 - 3) + 1.06193\ln(J)^4$ |
| | | 3% | $0.491(\bar{I}_1 - 3) + 1.00917(\bar{I}_2 - 3) + 1.21597\ln(J)^4$ |
| | | 5% | $0.463(\bar{I}_1 - 3) + 1.06173(\bar{I}_2 - 3) + 1.65483\ln(J)^4$ |
| **YH** | True | | $0.500\,(\bar{I}_1 - 3) + 0.700\,(\bar{I}_1 - 3)^2 + 0.300\,(\bar{I}_1 - 3)^3 + 1.500\ln(J)^2$ |
| | Noise level | 0% | $0.49085(\bar{I}_1 - 3) + 0.71488(\bar{I}_1 - 3)^2 + 0.29049(\bar{I}_1 - 3)^3 + 1.49851\ln(J)^2$ |
| | | 1% | $0.49440(\bar{I}_1 - 3) + 0.69077(\bar{I}_1 - 3)^2 + 0.29981(\bar{I}_1 - 3)^3 + 1.50495\ln(J)^2$ |
| | | 3% | $0.47797(\bar{I}_1 - 3) + 0.68731(\bar{I}_1 - 3)^2 + 0.32673(\bar{I}_1 - 3)^3 + 1.50140\ln(J)^2$ |
| | | 5% | $0.44767(\bar{I}_1 - 3) + 0.84085(\bar{I}_1 - 3)^2 + 0.22062(\bar{I}_1 - 3)^3 + 1.44813\ln(J)^2$ |
| **HN** | True | | $0.500\,(\bar{I}_1 - 3) + 1.000\left(\bar{I}_2^{3/2} - 3^{3/2}\right)^2 + 1.500(J - 1 - \ln(J))$ |
| | Noise level | 0% | $0.49242(\bar{I}_1 - 3) + 0.99649\left(\bar{I}_2^{3/2} - 3^{3/2}\right)^2 + 1.50762(J - 1 - \ln(J))$ |
| | | 1% | $0.48503(\bar{I}_1 - 3) + 1.00009\left(\bar{I}_2^{3/2} - 3^{3/2}\right)^2 + 1.52545(J - 1 - \ln(J))$ |
| | | 3% | $0.48503(\bar{I}_1 - 3) + 6.86699(\bar{I}_2 - 3)^2 + 1.15188(\bar{I}_2 - 3)^3 + 0.77310\ln(J)^2$ |
| | | 5% | $0.49749(\bar{I}_1 - 3) + 1.00749\left(\bar{I}_2^{3/2} - 3^{3/2}\right)^2\ 0.77277\ln(J)^2$ |

**Fig. 5** illustrates the evolution of the loss function during the training process for the YH model under varying noise levels. In Stage 1, both the constitutive parameters and neural network parameters are simultaneously optimized using the Adam optimizer, resulting in a gradual reduction in the loss function. In Stage 2, the parameters are updated alternately: the constitutive parameters are refined through sparse regression, while the neural network parameters are optimized using the L-BFGS method. During the training process, the loss function converges, with occasional peaks corresponding to updates in the constitutive parameters. These peaks, caused by sparse regression, are followed by rapid decreases as the parameters are further optimized After approximately five iterations of sparse regression, further decreases in the loss function are minimal, indicating that the model has effectively converged. Notably, higher noise levels yield increased final loss values, which underscores the framework's sensitivity to data quality. Additional details on the loss evolution during training, as well as a comparison between Stage 1 only and the complete two-stage optimization, are provided in **Supplementary Note 2.**

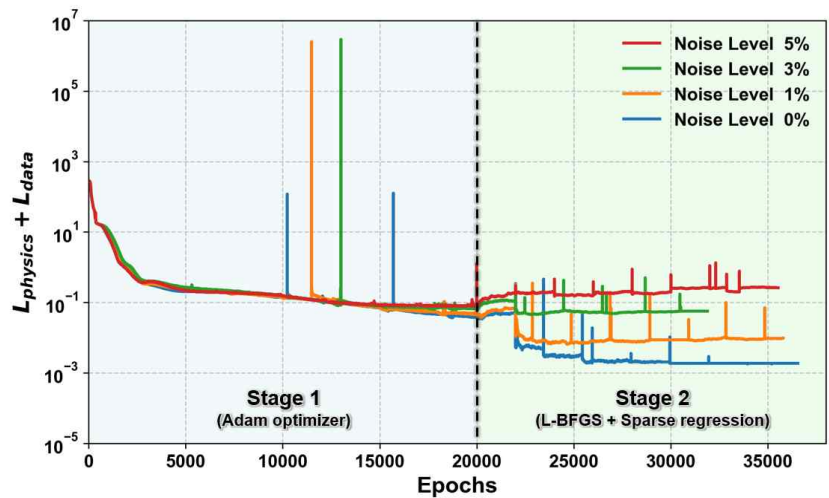

**Fig. 5.** Loss history during two-stage training of the benchmark model YH under different noise levels.

**Fig. 6** illustrates the progression of constitutive parameter optimization during the two-stage training process for benchmark models (**Fig. 6(a)**). The y-axis represents the index of the 57 constitutive parameters $\boldsymbol{\theta_W}$. A value of zero or almost zero less than $10^{-6}$ for a given index indicates that the corresponding strain energy density candidate is not selected in the final constitutive model. In Stage 1, both the constitutive parameters and neural network parameters are simultaneously optimized using the Adam optimizer. During this stage, multiple strain energy density candidates contribute to the material behavior, as reflected by several non-zero components. In Stage 2, sparse regression is applied, yielding a parsimonious set of constitutive parameters. Most components approach zero, leaving only a select few with non-zero values, thereby enabling precise discovery of the ground truth model. The constitutive parameter evolution for all eight benchmark models throughout the training process is presented in **Fig. 6 (b)**. These results demonstrate the framework's robustness in accurately refining and identifying the correct parameters across various benchmarks.

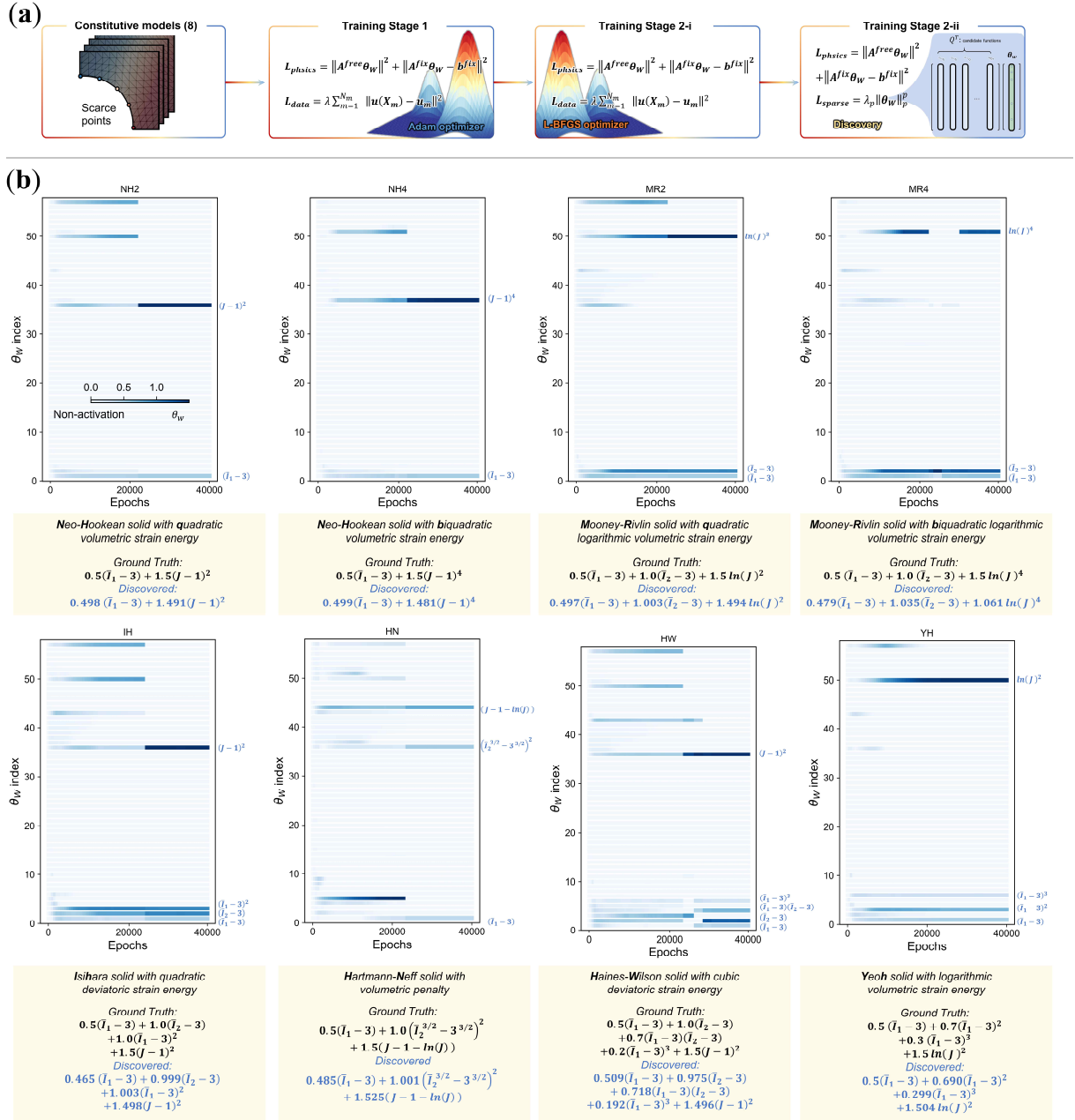

**Fig. 6. Evolution of Constitutive Parameters During the Two-Stage Training Process for Eight Benchmark Models at 1% Noise Level**. (a) Overview of the two-stage training strategy for optimizing constitutive parameters. (b) History of the 57 constitutive parameters ($\boldsymbol{\theta_W}$) values for eight benchmark models

Finally, sparse measured displacement data were employed to reconstruct the full-field displacement using the proposed PINN framework while concurrently identifying the underlying constitutive model. The reconstruction results shown in **Fig. 7** and **Fig. 8** correspond to the MR2 model at a 1% noise level. **Fig. 7** compares the reconstructed displacement fields with the ground-truth data from noise-free finite element analysis (FEA), with **Fig. 7(a)** illustrating the x-direction displacement predictions and **Fig. 7(b)** depicting the y-direction displacement predictions. The minimal differences between the reconstructed fields and the FEA results demonstrate the high accuracy and reliability of the proposed approach. Furthermore, the reconstructed displacement field was used to calculate the three primary invariants of the deformation gradient, as presented in **Fig. 8**, where the invariant fields—representing $\bar{I}_1$, $\bar{I}_2$, and $J$—exhibit good agreement with the FEA data, even in regions undergoing significant deformation (e.g., near the central hole). Notably, the two-stage training strategy—comprising an initial optimization with the Adam optimizer followed by refinement via L-BFGS combined with sparse regression—yields substantially lower errors compared to using only Stage 1 training. Overall, these results validate the proposed PINN framework as a robust method for accurately capturing material behavior and reliably discovering constitutive models under data-limited conditions.

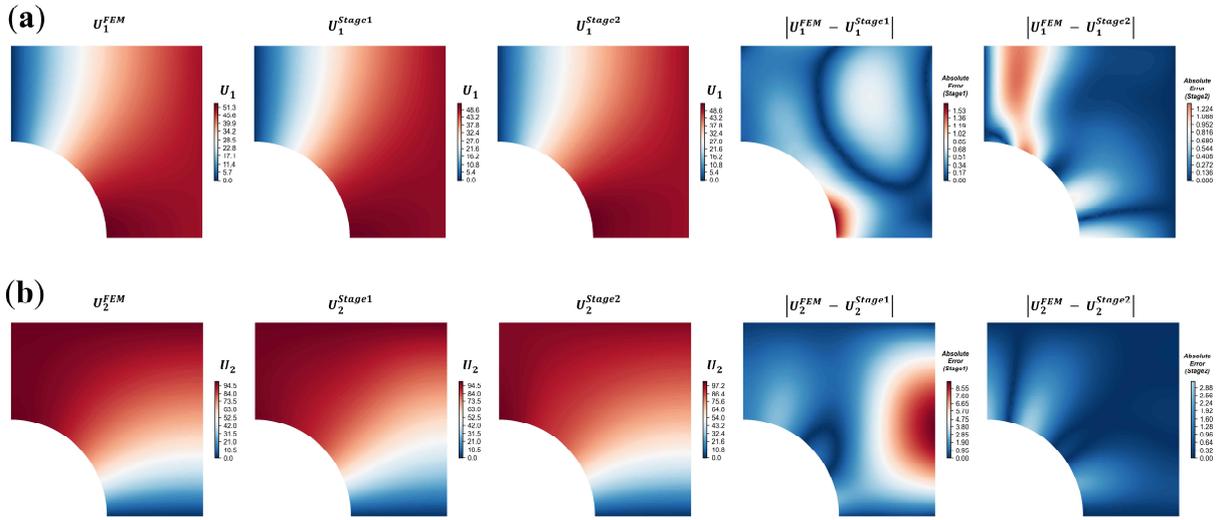

**Fig. 7. Reconstructed x- and y-direction displacement fields and corresponding error distributions comparing the proposed PINN predictions to the ground-truth FEA results for the MR2 model at 1% noise.** (a) x-direction component. (b) y-direction component.

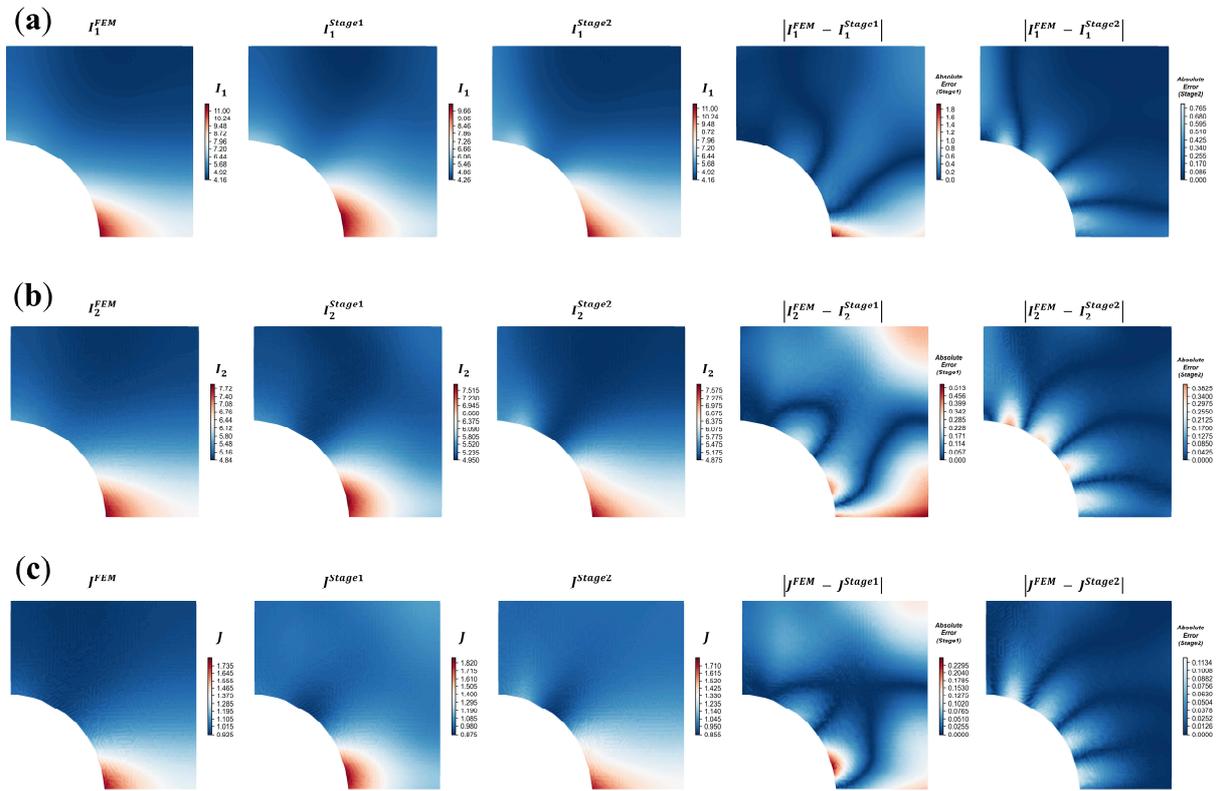

**Fig. 8. Reconstructed invariant fields and corresponding error distributions comparing the proposed PINN predictions to the ground-truth FEA results for the MR2 model at 1% noise.** (a) first invariant ($\bar{I}_1$), (b) second invariant ($\bar{I}_2$), and (c) third invariant ($J$).

## 5. Conclusion

In this study, we presented a PINN framework for discovering the strain energy density of hyperelastic materials using limited measurement data. The problem was formulated as an inverse problem with the dual objective of identifying the constitutive model parameters and reconstructing the displacement field based on a set of candidate strain energy density functions and the force equilibrium equation. To tackle this, we employed a Physics-Informed Neural Network (PINN) combined with finite element discretization and a weak-form loss formulation. A two-stage training strategy was introduced to jointly optimize neural network parameters and model coefficients.

The proposed approach was validated using eight benchmark hyperelastic models, demonstrating its capability to accurately recover the reference constitutive behavior. Moreover, the robustness of the framework was evaluated under varying levels of Gaussian noise in the measurement data, with the method successfully identifying the constitutive model even at noise levels as high as 3%, thereby highlighting its resilience to noisy experimental conditions. Despite the limited data, the framework exhibited notable performance, confirming its practicality in scenarios where high-quality measurement data are scarce. Overall, the approach offers a robust and data-efficient solution for constitutive model discovery in challenging experimental environments. By coupling PINN with efficient optimization strategies, this study demonstrates the capability to precisely identify constitutive models under complex conditions, while also outlining promising avenues for future research in material behavior modeling, particularly in scenarios where data are scarce and noisy.

**Declaration of Competing Interest**

The authors declare that they have no known competing financial interests or personal relationships that could have appeared to influence the work reported in this paper

**Data availability**

Data will be made available on request.

**Acknowledgements**

This work was supported by the National Research Foundation of Korea (NRF) grants funded by the Korean government (MSIT) (Nos. RS-2023-00222166 and RS-2023-00247245), and by a grant from the Ministry of Food and Drug Safety (No. RS-2023-00215667).

# Supplementary Information

# Physics-Informed Neural Network-Based Discovery of Hyperelastic Constitutive Models from Extremely Scarce Data

Hyeonbin Moon [1†], Donggeun Park [1†], Hanbin Cho [1], Hong-Kyun Noh [2], Jae hyuk Lim [2*] and Seunghwa Ryu [1*]

[1]Department of Mechanical Engineering, Korea Advanced Institute of Science and Technology (KAIST), 291 Daehak-ro, Yuseong-gu, Daejeon 34141, Republic of Korea

[2] Department of Mechanical Engineering, Kyung Hee University, 1732 Deogyeong-daero, Giheung-gu, Yongin-si, Gyeonggi-do 17104, Republic of Korea

[†] Hyeonbin Moon and Donggeun Park contributed equally to this work.

[*]Corresponding authors: ryush@kaist.ac.kr (Seunghwa Ryu), jaehyuklim@khu.ac.kr (Jae Hyuk Lim)

**Supplementary Note 1. Strain energy density candidates**

The following presents the complete set of 57 candidate functions considered in this study for constructing the strain energy density function. The selected functions include commonly used forms for isochoric and volumetric responses, such as those from the generalized Mooney–Rivlin and Hartmann–Neff models. While this set is sufficient for capturing a wide range of nonlinear behaviors, the formulation allows for straightforward extension by incorporating additional candidate terms as needed.

(1) 35 Candidates inspired by the generalized Mooney–Rivlin model

$i = 1$                       $(\bar{I}_1 - 3), \ (\bar{I}_2 - 3)$

$i = 2$                $(\bar{I}_1 - 3)^2, \ (\bar{I}_1 - 3)(\bar{I}_2 - 3), \ (\bar{I}_2 - 3)^2$

$i = 3$      $(\bar{I}_1 - 3)^3, \ (\bar{I}_1 - 3)^2(\bar{I}_2 - 3), \ (\bar{I}_1 - 3)(\bar{I}_2 - 3)^2, \ (\bar{I}_2 - 3)^2$

$\vdots$                                       $\vdots$

$i = 7$    $(\bar{I}_1 - 3)^7, \ (\bar{I}_1 - 3)^6(\bar{I}_2 - 3), \ (\bar{I}_1 - 3)^5(\bar{I}_2 - 3)^2, \ \cdots, \ (\bar{I}_1 - 3)(\bar{I}_2 - 3)^6, \ (\bar{I}_2 - 3)^7$

(2) 7 Candidates inspired by the Hartmann-Neff model

$$\left(\bar{I}_2^{3/2} - 3^{3/2}\right), \ \left(\bar{I}_2^{3/2} - 3^{2/3}\right)^2, \ \left(\bar{I}_2^{3/2} - 3^{2/3}\right)^3,$$
$$\left(\bar{I}_2^{3/2} - 3^{2/3}\right)^4, \ \left(\bar{I}_2^{3/2} - 3^{2/3}\right)^5, \ \left(\bar{I}_2^{3/2} - 3^{2/3}\right)^6, \ \left(\bar{I}_2^{3/2} - 3^{2/3}\right)^7$$

(3) 15 Candidates for volumetric strain energy density

$$(J - 1)^2, \ (J - 1)^4, \ (J - 1)^6, \ (J - 1)^8, (J - 1)^{10}, \ (J - 1)^{12}, \ (J - 1)^{14}$$
$$\hbar(J)^2, \ \hbar(J)^4, \ \hbar(J)^6, \ \hbar(J)^8, \hbar(J)^{10}, \ \hbar(J)^{12}, \ \hbar(J)^{14}$$
$$J - 1 - \hbar(J)$$

**Supplementary Note 2. Effect of Two-Stage Training on Loss Convergence**

  **Fig. S1** compares the performance of two-stage training against Stage 1–only training for the YH benchmark model across varying noise levels. The x-axis corresponds to the physics loss, reflecting the model's compliance to governing equations, and the y-axis indicates the data loss, measuring the discrepancy between predicted and measured displacements. In **Fig. S1(a)**, which employs the full two-stage strategy, both losses converge to relatively lower values under lower noise conditions, signifying that the additional training stage effectively balances physical consistency with measured data. By contrast, **Fig. S1(b)** illustrates that relying solely on Stage 1—without the subsequent refinement—yields higher losses for the same noise levels, thereby underscoring the importance of the second stage in achieving robust convergence. The red markers in both cases indicate the final loss values achieved by the two-stage method, consistently revealing a loss gap compared to the Stage 1–only case.

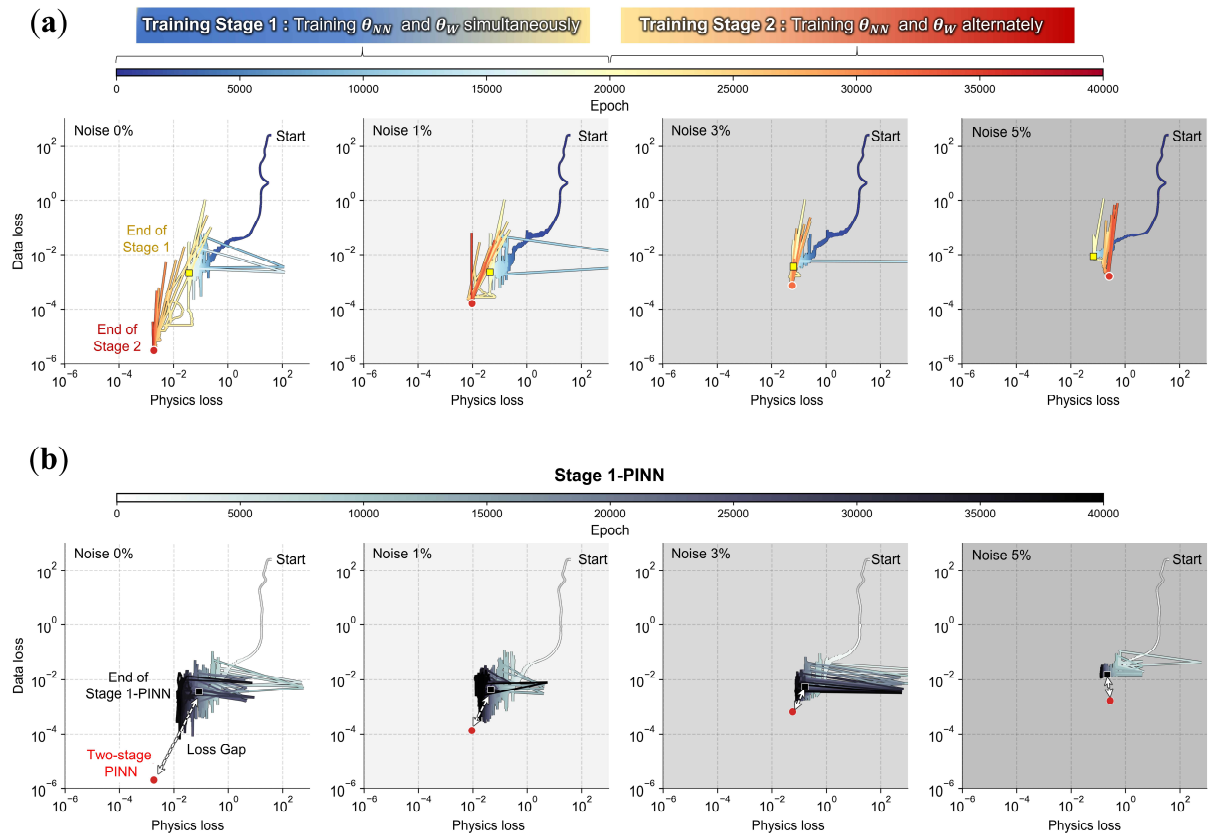

**Fig. S1. Loss history for the benchmark model YH at noise levels of 0%, 1%, 3%, and 5% under different training strategies. (**a) Loss history during two-stage PINN training. (b) Loss history when using only Stage 1 with the Adam optimizer.